\documentclass[12pt,prb,floatfix,showpacs,aps,citeautoscript]{revtex4-1}
\usepackage{hyperref,amsmath,graphicx,amssymb}
\usepackage[english]{babel}

\newcommand{\ket}[1]{|#1\rangle}
\newcommand{\ave}[1]{\langle#1\rangle}
\newcommand{\bra}[1]{\langle#1|}
\newcommand{\oforder}[1]{\mathcal{O}(#1)}
\newcommand{\mate}[3]{\left\langle#1|#2|#3\right\rangle}
\newcommand{\matens}[3]{\langle#1|#2|#3\rangle} 
\newcommand{\mtext}[2]{\hspace{#1cm}{\mathrm{#2}}\hspace{#1cm}}
\newcommand{\limitarrow}{\stackrel{\ell\rightarrow\infty}{\longrightarrow}}
\newcommand{\epst}{\tilde{\epsilon}}

\begin{document}
\title{Dynamical quantum phase transitions in the axial next-nearest-neighbour Ising chain}  
\date{\today}
\author{JN Kriel$^1$, C Karrasch$^{2,3}$, S Kehrein$^4$}
\affiliation{$^1$Institute of Theoretical Physics, University of Stellenbosch, Stellenbosch 7600, South Africa}
\affiliation{$^2$Department of Physics, University of California, Berkeley, CA 95720, USA}
\affiliation{$^3$Materials Sciences Division, Lawrence Berkeley National Laboratory, Berkeley, CA 94720, USA}
\affiliation{$^4$Institut f\"{u}r Theoretische Physik, Georg-August-Universit\"{a}t G\"{o}ttingen, D-37077 G\"{o}ttingen, Germany}
\begin{abstract}
We investigate sudden quenches across the critical point in the transverse field Ising chain with a perturbing non-integrable next-nearest-neighbour interaction. Expressions for the return (Loschmidt) amplitude and associated rate function are derived to linear order in the next-nearest-neighbour coupling. In the thermodynamic limit these quantities exhibit non-analytic behaviour at a set of critical times, a phenomenon referred to as a dynamical quantum phase transition. We quantify the effect of the integrability breaking perturbation on the location and shape of these non-analyticities. Our results agree with those of earlier numerical studies and offer further support for the assertion that the dynamical quantum phase transitions exhibited by this model are a generic feature of its post-quench dynamics and is robust with respect to the inclusion of non-integrable perturbations.
\end{abstract}
\maketitle
\section{Introduction}
Advances in the experimental manipulation of systems such as cold atomic gasses \cite{greiner_2002,kinoshita_2006} has allowed for the realisation of unitary time evolution in closed quantum systems \cite{polkovnikov_2011}. This has triggered much theoretical interest in non-equilibrium quantum dynamics, particularly in relation to the existence and characterisation of long-time stationary states. A typical scenario in this context is that of a quantum quench, in which a system is driven out of equilibrium by tuning a control parameter, typically an external field strength. In this paper our interest lies with the finite-time dynamics following a sudden quench, and the emergence of non-analytic behaviour in certain quantities in the thermodynamic limit. To set the scene, consider the return (Loschmidt) amplitude
\begin{equation}
	G(t)=\mate{\Psi_0}{e^{-iH t}}{\Psi_0}
\end{equation}
with $\ket{\Psi_0}$ the initial state and $H$ the Hamiltonian driving the post-quench dynamics. Heyl et al. \cite{heyl_2013} noted the formal similarity between $G(t)$ and the canonical partition function $Z(\beta)={\mathrm{tr}}(e^{-\beta H})$. As is well known from the Lee-Yang treatment of equilibrium phase transitions the non-analytic behaviour of the free energy density can be understood by analysing the Fischer zeros of $Z(\beta)$ in the complex temperature plane \cite{fisher_1965}. In this spirit Heyl et al. investigated the analytic behaviour of the boundary partition function $Z(z)=\mate{\Psi_0}{e^{-z H}}{\Psi_0}$ with $z\in\mathbb{C}$ for quenches in the transverse field Ising chain. It was found that in the thermodynamic limit, and for quenches between the paramagnetic and ferromagnetic phases, the zeros of $Z(z)$ coalesce into lines which intersect the time axis. This results in non-analytic behaviour in the rate function of the return probability $l(t)=\lim_{L\rightarrow\infty}-L^{-1}\log|G(t)|^2$ at a set of critical times $t^*_n$. At these times the system is said to exhibit a \emph{dynamical quantum phase transition}. Furthermore, these transitions were shown to impact on the behaviour of the experimentally relevant work distribution function, while the critical times themselves introduce a new quench-dependent time scale which enters in the dynamics of the order parameter. Aspects of this phenomenon have since been the focus of a number of studies \cite{gong_2013,fagotti_2013,heyl_2014,vajna_2014,andraschko_2014,hickey_2014}. In particular, Karrasch and Schuricht \cite{karrasch_2013} investigated the robustness of these phase transitions for quenches in two non-integrable spin models using the time-dependent density-matrix renormalisation group (tDMRG) algorithm. It was found that the dynamical phase transitions persist is the presence of non-integrable interactions, although the shape and location of the non-analyticities get modified in a non-trivial way.\\

In this paper we complement this study with analytic calculations for quenches in the transverse field Ising chain perturbed by a non-integrable next-nearest-neighbour (NNN) interaction. The Hamiltonian driving the dynamics is then the axial transverse next-nearest-neighbour Ising (ANNNI) model \cite{selke_1988,suzuki_2013}. To reliably describe the dynamics at longer times we implement the continuous unitary transformations (CUTs) approach to calculate the rate function of the return probability to linear order in the NNN coupling.\\

The paper is organised as follows. In section \ref{sectionIsing} we summarise some results from Refs. \onlinecite{silva_2008,heyl_2013} for quenches in the transverse field Ising chain. The CUTs diagonalization procedure is outlined in section \ref{sectionCUTs} and used in sections \ref{sectionTA} and \ref{sectionRF} for the perturbative calculation of the return probability and rate function for quenches to the ANNNI model. These results are benchmarked against tDMRG calculations in section \ref{sectionCompareDMRG}. In section \ref{sectionAnalyse} we analyse how the shape and location of the non-analyticies in the rate function are modified by the NNN interaction. Section \ref{sectionConclusion} concludes the paper. Some technical details of the calculations appear in the appendix.
\section{Quenches in the Transverse Field Ising Chain}
\label{sectionIsing}
\begin{figure}[ht]
\hspace{-0.5cm}
  \begin{minipage}[b]{0.47\linewidth} \centering \includegraphics[height=5.7cm]{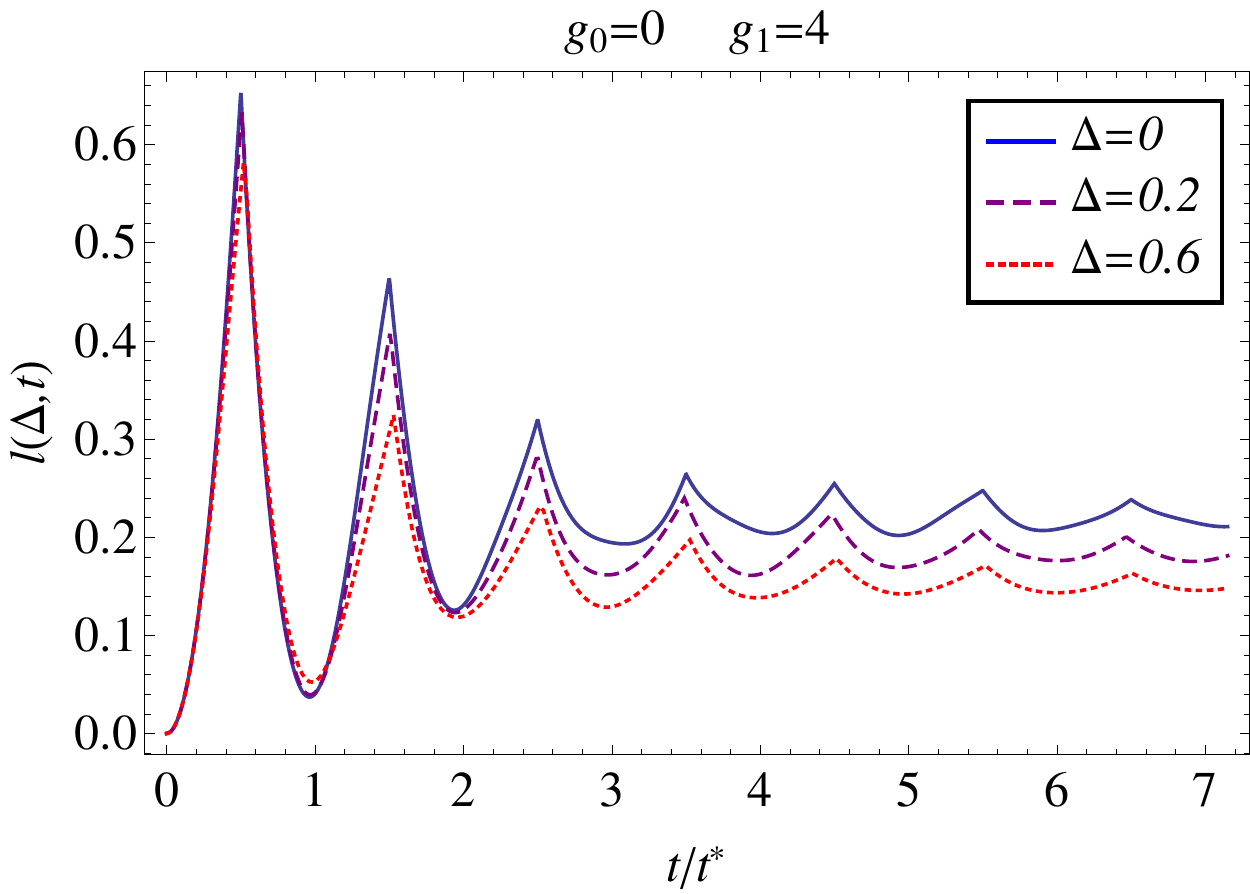} \end{minipage}
  \hspace{0.6cm}
  \begin{minipage}[b]{0.47\linewidth} \centering \includegraphics[height=5.7cm]{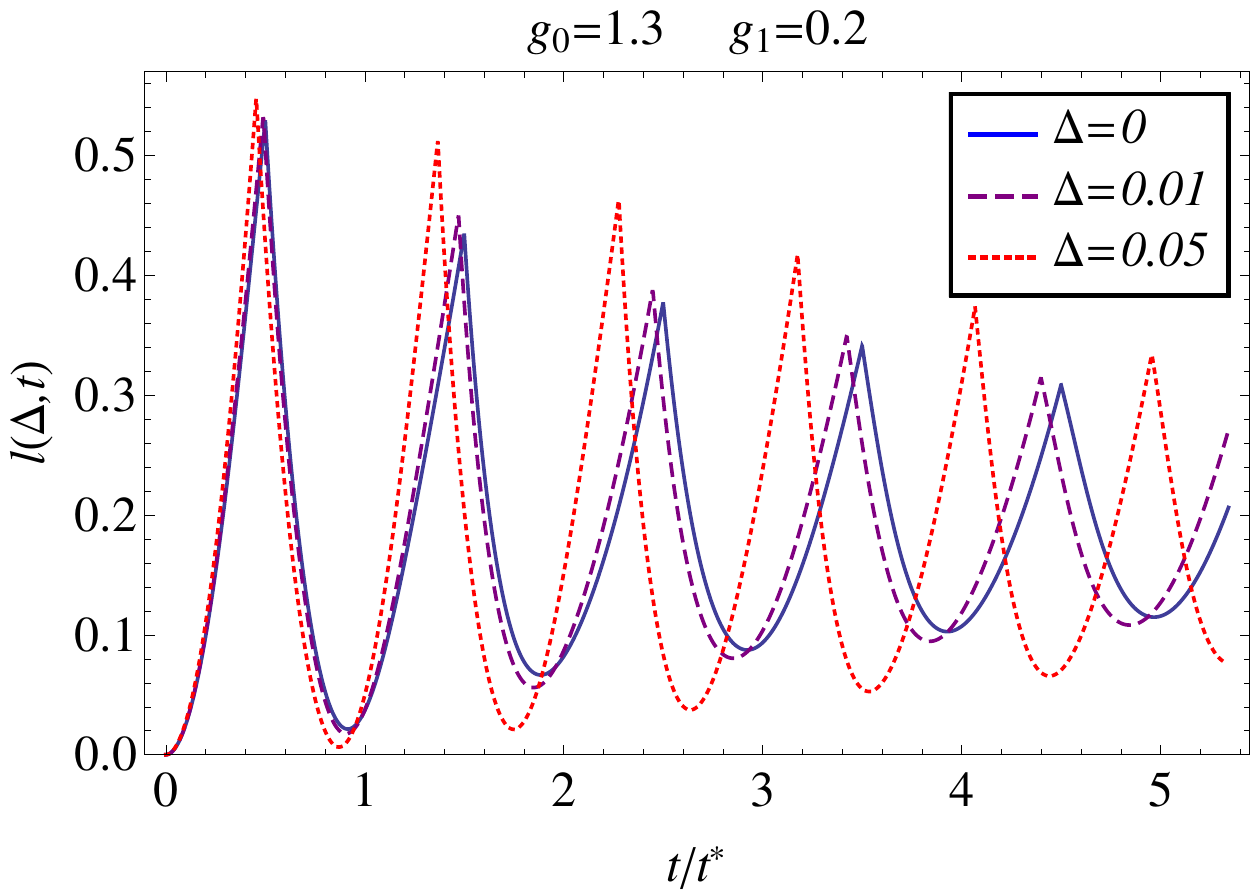} \end{minipage}
	\caption{The rate function $l(\Delta,t)$ for quenches from the FM to PM phase (left) and PM to FM phase (right). The $\Delta=0$ curve corresponds to \eqref{H0results}. Results for $\Delta>0$ were obtained using the tDMRG algorithm. See Section \ref{sectionCompareDMRG} and Ref. \onlinecite{karrasch_2013} for details.}
	\label{figureratefunctions}
\end{figure}

The one-dimensional transverse field Ising model is 
\begin{equation}
	H_0(g)=-\sum_{i=1}^L(\sigma_i^z\sigma_{i+1}^z+g\sigma_i^x)
	\label{H0spin}
\end{equation}
with periodic boundary condition $\sigma^z_{L+1}=\sigma^z_1$ and where $g$ denotes the transverse magnetic field strength. This model exhibits a quantum phase transition at $g=g_c=1$ from a ferromagnetic $(g<1)$ to a paramagnetic $(g>1)$ phase \cite{sachdev_2011}. It is exactly solvable through a combination of a Wigner-Jordan and Bogoliubov transformation which produces a description in terms of free fermions. The dynamics of this model following a quench in $g$ has been studied by a number authors \cite{heyl_2013,pollmann_2010,calabrese_2011,silva_2008,karrasch_2013} and we only summarise some basic results here. In a quantum quench experiment the system is prepared in the ground state of an initial Hamiltonian $H_0(g_0)$ and then allowed to evolve unitarily under the final Hamiltonian $H_0(g_1)$. Let $\{\eta^\dag_k,\eta_k\}$ and $\{\gamma^\dag_k,\gamma_k\}$ denote the fermionic species diagonalising $H_0(g_0)$ and $H_0(g_1)$ respectively. We have\cite{sachdev_2011,calabrese_2012}
\begin{equation}
\label{H0fermionic}
	H_0(g_0)=\sum_k \epsilon_k(g_0)[\eta^\dag_k\eta_k-1/2] \mtext{0.8}{and} H_0(g_1)=\sum_k \epsilon_k(g_1)[\gamma^\dag_k\gamma_k-1/2]
\end{equation}
where $\epsilon_k(g)=2\sqrt{(g-\cos k)^2+\sin^2k}$. The two species are related by $\eta_k=U_k\gamma_k+iV_k\gamma^\dag_{-k}$ where $U_k=\cos(\phi_k)$ and $V_k=\sin(\phi_k)$ with $\phi_k=\theta_k(g_1)-\theta_k(g_0)$ and $\tan(2\theta_k(g))=\sin k/(g-\cos k)$. The quantities of interest here are the return (Loschmidt) amplitude $G(t)={_\eta}{\matens{0}{e^{-itH_0(g_1)}}{0}}{_\eta}$ and the rate function of the return probability
\begin{equation}
	l(t)=-\lim_{L\rightarrow\infty}\frac{1}{L}\log|G(t)|^2.
	\label{ratefunction1}
\end{equation}
Here $\ket{0}_\eta$ is the $\eta$-vacuum and the ground state of $H_0(g_0)$. The latter is related to the $\gamma$-vacuum through
\begin{equation}
	\ket{0}_\eta=N^{-1}e^{-i\sum_{k>0}\Lambda_k\gamma^\dag_k\gamma^\dag_{-k}}\ket{0}_\gamma
	\label{etavacuum}
\end{equation}
with $N^2=\prod_{k>0}(1+\Lambda_k^2)$ and $\Lambda_k=V_k/U_k=\tan\phi_k$. It now follows that\cite{silva_2008}
\begin{equation}
	G(t)=\prod_{k>0}\left(U_k^2+V_k^2 e^{-2it\epsilon_k(g_1)}\right)\mtext{0.8}{and}l(t)=-2\int_0^\pi\frac{dk}{2\pi}\ln|U_k^2+V_k^2 e^{-2it\epsilon_k(g_1)}|.
	\label{H0results}
\end{equation}
For quenches across the phase transition this quantity exhibits non-analytic behaviour in the form of cusps which appear periodically at the critical times 
\begin{equation}
\label{H0criticaltimes}
	t^*_n=t^*(n+1/2),\ \ \ n=0,1,2,\ldots
\end{equation}
with $t^*=\pi/\epsilon_{k^*}(g_1)$ and $\cos k^*=(1+g_0 g_1)/(g_0+g_1)$. These non-analyticities are a result of $G(t)$ factorising into contributions from the various $k$-modes together with the existence of a particular mode $k^*$ which satisfies $U^2_{k^*}=V^2_{k^*}$, and for which the argument of the logarithm in \eqref{H0results} vanishes at $t=t^*_n$.  This is illustrated in Figure \ref{figureratefunctions}. It is clear that integrable perturbations that still allow for a free-fermion description will not fundamentally alter this picture. However, it is less obvious that this phenomenon persists in the presence of non-integrable interactions. \\

A final important point remains to be addressed. After applying the Wigner-Jordan transformation to the spin Hamiltonian in \eqref{H0spin} the fermionic Fock space is found to factorize into sectors with even and odd particle numbers. In the even (Neveu-Schwarz) sector it is natural to impose anti-periodic boundary conditions on the fermions, and this leads to a quantisation of the momentum in half-integer multiples of $2\pi/L$. In the odd (Ramond) particle number sector we enforce periodic boundary conditions leading to momentum quantisation in integer multiples of $2\pi/L$. At finite $L$, and for all $g$, the system's true ground state lies in the even sector \cite{calabrese_2012}. In the ferromagnetic phase this state is a superposition of symmetry broken polarised states. In the thermodynamic limit the ground states of the odd and even sectors become degenerate and one recovers the two polarised ferromagnetic ground states. We emphasise that the expressions in \eqref{H0results} are applicable only to quenches starting from the mixed ground state of the even sector. We focus on this case in what follows.
\section{Quenches in the ANNNI Model}
\label{sectionANNNI}
We now turn to quenches which involve tuning $g$ across the phase transition while simultaneously switching on a non-integrable next-nearest-neighbour interaction. The initial Hamiltonian remains $H_0(g_0)$ while the time evolution is now generated by the ANNNI Hamiltonian \cite{selke_1988,suzuki_2013}
\begin{equation}
\label{fullH}
	H(g_1,\Delta)=-\sum_{i=1}^L(\sigma_i^z\sigma_{i+1}^z+g_1\sigma_i^x+\Delta\sigma_i^z\sigma_{i+2}^z)=H_0(g_1)+H_1(\Delta).
\end{equation}
The behaviour of the rate function $l(\Delta,t)$ following quenches in this model have previously been studied using the tDMRG algorithm in Ref. \onlinecite{karrasch_2013}. Results appear in Figure \ref{figureratefunctions} for two quenches and various values of $\Delta$. The shape and locations of the cusps appear to depend on the NNN coupling in a regular way, even at long times and for a range of coupling strengths. Even strong coupling should therefore not fundamentally alter the nature of these non-analytic structures, provided, of course, that the system is not driven into a different phase. This suggests that the qualitative effect of the NNN interaction can be captured well within a perturbative framework.\\

Our goal in what follows is to calculate the linear order correction to the rate function $l(\Delta,t)$ due to this perturbing interaction. For this purpose standard time-dependent perturbation theory is not sufficient, as it produces secular terms which grow linearly in time, leading to an eventual breakdown in the perturbative approximation \cite{hackl_2009}. To overcome this problem we make use of the continuous unitary transformations (CUTs) approach \cite{wegner_1994,kehrein_2006}. This technique has been applied successfully to a variety of non-equilibrium problems \cite{kehrein_2005,hackl_2008,moeckel_2008,moeckel_2009,moeckel_2010,essler_2014}. The $g_1$ and $\Delta$ arguments of $H_{0,1}$ are suppressed in what follows.
\subsection{Diagonalisation via CUTs}
\label{sectionCUTs}
In the CUTs approach a sequence of infinitesimal unitary transformations is used to bring the Hamiltonian into an energy diagonal form.  Following this, states and observables may be evolved in time using this diagonalised Hamiltonian without the risk of producing secular terms. The evolution of the Hamiltonian under this sequence of transformations is parametrised by a flow parameter $\ell$ and governed by the equation  
\begin{equation}
	\frac{d H(\ell)}{d\ell}=[\Gamma(\ell),H(\ell)]
	\label{floweq}
\end{equation}
where $\Gamma(\ell)$ is an antihermitian generator. The post-quench Hamiltonian $H=H_0+H_1$ provides the initial condition at $\ell=0$, i.e. $H(0)=H$. At finite $\ell$ this Hamiltonian is unitarily transformed into $H(\ell)=U(\ell)H(0)U^\dag(\ell)$ where $U(\ell)$ satisfies $d U(\ell)/d\ell=\Gamma(\ell)U(\ell)$ and $U(0)=I$. By choosing the generator $\Gamma(\ell)$ appropriately we can ensure that the flow converges to a fixed point $H(\infty)$ which is diagonal in the eigenbasis of a chosen non-interacting Hamiltonian. For the latter we take simply $H_0$ and set $\Gamma(\ell)=[H_0,H(\ell)]$ which is known to produce a fixed point for which $[H_0,H(\infty)]=0$, i.e. which is ``energy diagonal'' with respect to the unperturbed Hamiltonian $H_0$. Transforming to a description in terms of the $\gamma$-fermions of \eqref{H0fermionic} we have, as before, that $H_0=\sum_k \epsilon_k(g_1)[\gamma^\dag_k\gamma_k-1/2]$ while the interaction term reads
\begin{align}
	H_1=A&+\sum_k B(k)\gamma_k^\dag\gamma_k+\sum_k \left[C(k)\gamma_k^\dag\gamma_{-k}^\dag+h.c.\right]+\sum_\mathbf{k}D(\mathbf{k})\gamma_{k_1}^\dag\gamma_{k_2}^\dag\gamma_{k_3}\gamma_{k_4}\nonumber\\
	&+\sum_\mathbf{k}\left[E(\mathbf{k})\gamma_{k_1}^\dag\gamma_{k_2}^\dag\gamma^\dag_{k_3}\gamma_{k_4}+h.c.\right]+\sum_\mathbf{k}\left[F(\mathbf{k})\gamma_{k_1}^\dag\gamma_{k_2}^\dag\gamma^\dag_{k_3}\gamma^\dag_{k_4}+h.c.\right].
	\label{H1fermionic}
\end{align}
Expressions for the various coefficients appear in the appendix. To linear order in $\Delta$ the flow described by \eqref{floweq} preserves the form of the original Hamiltonian $H=H_0+H_1$ with only the coefficients of the energy off-diagonal terms in $H_1$ evolving as
\begin{alignat}{4}
	C(k,\ell) &= \exp[-(2\epsilon_k)^2\ell]C(k) &\limitarrow\quad& 0 \\
	D(\mathbf{k},\ell) &= \exp[-E_D(\mathbf{k})^2\ell]D(\mathbf{k})\quad &\limitarrow\quad& \delta_{E_D(\mathbf{k}),0}D(\mathbf{k})\label{Dflow}\\
	E(\mathbf{k},\ell) &= \exp[-E_E(\mathbf{k})^2\ell]E(\mathbf{k}) &\limitarrow\quad& \delta_{E_E(\mathbf{k}),0}E(\mathbf{k})=0\\
	F(\mathbf{k},\ell) &= \exp[-E_F(\mathbf{k})^2\ell]F(\mathbf{k}) &\limitarrow\quad& \delta_{E_F(\mathbf{k}),0}F(\mathbf{k})=0
\end{alignat}
where $E_D(\mathbf{k})=\epsilon_{k_1}+\epsilon_{k_2}-\epsilon_{k_3}-\epsilon_{k_4}$, $E_E(\mathbf{k})=\epsilon_{k_1}+\epsilon_{k_2}+\epsilon_{k_3}-\epsilon_{k_4}$ and $E_F(\mathbf{k})=\epsilon_{k_1}+\epsilon_{k_2}+\epsilon_{k_3}+\epsilon_{k_4}$. This can be verified by substituting $H(\ell)$ into \eqref{floweq} and using, for example, 
\begin{equation}
	[[H_0,\gamma_{k_1}^\dag\gamma_{k_2}^\dag\gamma_{k_3}\gamma_{k_4}],H_0]=-E_D(\mathbf{k})^2 \gamma_{k_1}^\dag\gamma_{k_2}^\dag\gamma_{k_3}\gamma_{k_4}
\end{equation}
to check \eqref{Dflow}, and similar for the other coefficients. These are the only type of double commutators relevant at linear order since the coefficients of $H_1$ are already of order $\oforder{\Delta}$. Here and in what follows we abbreviate $\epsilon_k=\epsilon_k(g_1)$ and assume that $g_1\neq 1$, which ensures that $\epsilon_k>0$. As $\ell\rightarrow\infty$ the energy off-diagonal terms therefore decay exponentially, leaving only terms which commute with $H_0$. The combined constraints of momentum and energy conservation are responsible for $E(\mathbf{k},\infty)$ and $F(\mathbf{k},\infty)$ vanishing. Up to an additive constant the final Hamiltonian is
\begin{align}
	H(\infty)&=\sum_k (\epsilon_k+B(k))\gamma_k^\dag\gamma_k+\sum_\mathbf{k}\delta_{E_D(\mathbf{k}),0}D(\mathbf{k})\gamma_{k_1}^\dag\gamma_{k_2}^\dag\gamma_{k_3}\gamma_{k_4}\\
	&\approx\sum_k \tilde{\epsilon}_k\gamma_k^\dag\gamma_k+\sum_{k,k'}D_{k,k'}\gamma_{k'}^\dag\gamma_{k'}\gamma_{k}^\dag\gamma_{k}=\tilde{H}_0+\tilde{H}_1
\end{align}
with $\epst_k=\epsilon_k+B(k)$ and $D_{k,k'}=D(k,k',k',k)-D(k,k',k,k')$. The expression above is exact for odd $L$, while for even $L$ there are $\oforder{L}$ additional terms of the form $\gamma_{k}^\dag\gamma^\dag_{\pi-k}\gamma_{-k}\gamma_{k-\pi}$ which also enter in $\tilde{H}_1$. However, since $D(\mathbf{k})=\oforder{L^{-1}}$ these terms do not contribute extensively to $H(\infty)$ and may be neglected in the thermodynamic limit. The transformation relating $H(\infty)$ to $H=H(0)$ is given by the $\ell$-ordered exponential
\begin{equation}
	U(\infty)=\mathcal{T}_\ell\left\{\exp\left[\int_0^\infty d\ell\,\Gamma(\ell)\right]\right\}.
\end{equation}
All the energy off-diagonal terms in $H(\ell)$ are at least linear in $\Delta$ and so $\Gamma(\ell)=[H_0,H(\ell)]=\oforder{\Delta}$. It is therefore permissible to neglect the ordering prescription above when working to linear order and approximate the transformation by
\begin{equation}
	U(\infty)\approx\exp\left[\int_0^\infty d\ell\,\Gamma(\ell)\right]=\exp[J]
	\label{Uapprox}
\end{equation}
where
\begin{align}
	J&=\sum_k \left[\bar{C}(k)\gamma_k^\dag\gamma_{-k}^\dag-h.c.\right]+\sum_\mathbf{k}{}^{\boldsymbol{'}}\bar{D}(\mathbf{k})\gamma_{k_1}^\dag\gamma_{k_2}^\dag\gamma_{k_3}\gamma_{k_4}\nonumber\\
	&+\sum_\mathbf{k}\left[\bar{E}(\mathbf{k})\gamma_{k_1}^\dag\gamma_{k_2}^\dag\gamma^\dag_{k_3}\gamma_{k_4}-h.c.\right]+\sum_\mathbf{k}\left[\bar{F}(\mathbf{k})\gamma_{k_1}^\dag\gamma_{k_2}^\dag\gamma^\dag_{k_3}\gamma^\dag_{k_4}-h.c.\right]
	\label{Jexpression}
\end{align}
with $\bar{X}(\mathbf{k})=X(\mathbf{k})/E_X(\mathbf{k})$ for $X=C,D,E,F$. In the primed summation those terms for which $E_D(\mathbf{k})=0$ are excluded. 
\subsection{Transition Amplitude}
\label{sectionTA}
Combining \eqref{Uapprox} with the identity $H=U^\dag(\infty)H(\infty)U(\infty)$ allows the transition amplitude to be approximated as
\begin{align}
	G(t)&={_\eta}{\matens{0}{e^{-itH}}{0}}{_\eta}={_\eta}{\matens{0}{U^\dag(\infty)e^{-itH(\infty)}U(\infty)}{0}}{_\eta}\\
	&\approx{_\eta}{\matens{0}{e^{-J}e^{-it\tilde{H}_1}e^{-it\tilde{H}_0}e^J}{0}}{_\eta}.
	\label{transitionamplitude}
\end{align}
From here there are several possible routes which lead to expressions for $l(t)$ which are equivalent up to linear order in $\Delta$. We will proceed in the spirit of the CUTs approach and avoid the truncation of exponential power series based on perturbative approximations, as this may well reintroduce secular terms. As a first step we rewrite the $e^{-it\tilde{H}_1}$ factor in \eqref{transitionamplitude} as
\begin{equation}
	e^{-it\tilde{H}_1}=\prod_{k,k'}\left[1+(e^{-itD_{k,k'}}-1)\gamma^\dag_{k'}\gamma_{k'}\gamma^\dag_{k}\gamma_{k}\right]
	\label{exph1factor}
\end{equation}
and then approximate $G(t)$ by
	\begin{equation}
		G(t)\approx\prod_{k,k'}\left[1+(e^{-itD_{k,k'}}-1)\ave{\gamma^\dag_{k'}\gamma_{k'}}\ave{\gamma^\dag_{k}\gamma_{k}}\right] {_\eta}{\matens{0}{e^{-J}e^{-it\tilde{H}_0}e^J}{0}}{_\eta}
		\label{gfactorise}
	\end{equation}
where 
\begin{equation}
	\ave{\hat{O}}=\frac{{_\eta}{\matens{0}{\hat{O}e^{-it \tilde{H}_0}}{0}}{_\eta}}{{_\eta}{\matens{0}{e^{-it \tilde{H}_0}}{0}}{_\eta}}.
\end{equation}
This approximation relies on two observations. First, since both $(e^{-itD_{k,k'}}-1)$ and $J$ are of order $\oforder{\Delta}$ the $e^{\pm J}$ factors may be neglected when calculating the matrix elements of the number operators as this can only introduce higher order corrections. Secondly, we note the factorisation property 
\begin{equation}
	\ave{\prod_i \gamma^\dag_{k_i}\gamma_{k_i}}=\prod_i \ave{\gamma^\dag_{k_i}\gamma_{k_i}}
	\label{factorise}
\end{equation}
which is a consequence of \eqref{etavacuum} and holds whenever $k_i\neq\pm k_j$ for all $i\neq j$. Expanding the right hand side of \eqref{exph1factor} makes it clear that at any finite order in $\Delta$ the number of terms for which this factorisation fails is suppressed by a factor of $1/L$ relative to the number of completely factorizable terms. This justifies the factorisation of the matrix elements in \eqref{gfactorise} in the thermodynamic limit. From \eqref{etavacuum} the factors on the right of \eqref{factorise} are found to be 
\begin{equation}
	\ave{\gamma^\dag_k \gamma_k}=\Lambda_k^2 Q_k\mtext{1}{with}Q_k=(\Lambda_k^2+e^{2it\epst_k})^{-1}.
	\label{numberopev}
\end{equation}

What remains is to calculate the matrix element on the right of \eqref{gfactorise}. To leading order in $\Delta$ in the arguments of the exponentials it holds that
\begin{equation}
	{_\eta}{\matens{0}{e^{-J}e^{-it\tilde{H}_0}e^J}{0}}{_\eta}={_\eta}{\matens{0}{e^{e^{-it \tilde{H}_0} J e^{+it \tilde{H}_0}-J}e^{-it \tilde{H}_0}}{0}}{_\eta}
	\label{Gfactor}
\end{equation}
where the Baker-Campbell-Hausdorff formula has been used to combine the two exponentials involving $J$ according to $e^{\Delta \hat{A}}e^{\Delta \hat{B}}=e^{\Delta \hat{A}+\Delta \hat{B}+\mathcal{O}(\Delta^2)}$. We now introduce
\begin{equation}
	{\textstyle \mathcal{T}=\exp\left[-i\sum_{k>0} \Lambda_k e^{-2it\epst_k}\gamma^\dag_{k}\gamma^\dag_{-k}\right]\exp\left[i\sum_{k>0} Q_k\Lambda_k e^{2it\epst_k}\gamma_{k}\gamma_{-k}\right]}
\end{equation}
with $Q_k$ as in \eqref{numberopev}, and set $\mathcal{A}=e^{-it \tilde{H}_0} J e^{+it \tilde{H}_0}-J$. Applying \eqref{etavacuum} to the right of \eqref{Gfactor} produces, after some straightforward manipulations, 
\begin{equation}
	{\textstyle {_\eta}{\matens{0}{e^{\mathcal{A}}e^{-it \tilde{H}_0}}{0}}{_\eta}={_\gamma}{\matens{0}{e^{\mathcal{T}^{-1}\mathcal{A}\mathcal{T}}}{0}}{_\gamma}\,\prod_{k>0}\left(U_k^2+V_k^2 e^{-2it\tilde{\epsilon}_k}\right).}
\label{Gfactor2}
\end{equation}
The transformation $\mathcal{T}$ acts on the $\gamma_k^{(\dag)}$ operators in $\mathcal{A}$ according to
\begin{align}
	\bar{\gamma}_k&=\mathcal{T}^{-1}\gamma_k\mathcal{T}=Q_k e^{2it\epst_k} \gamma_k-i\Lambda_k e^{-2it\epst_k}\gamma^\dag_{-k}\\
	\bar{\gamma}^\dag_k&=\mathcal{T}^{-1}\gamma^\dag_k\mathcal{T}=\gamma^\dag_k+i\Lambda_k Q_k e^{2it\epst_k}\gamma_{-k}.
\end{align}
Through normal ordering $\mathcal{T}^{-1}\mathcal{A}\mathcal{T}$ can be brought into the form $\mathcal{T}^{-1}\mathcal{A}\mathcal{T}=\mathcal{A}_R+\mathcal{A}_L+\mathcal{A}_C$ where $\mathcal{A}_{R,L}$ are operators satisfying $\mathcal{A}_R\ket{0}_\gamma=0$ and ${_\gamma}\bra{0}\mathcal{A}_L=0$ and with $\mathcal{A}_C={_\gamma}\matens{0}{\mathcal{T}^{-1}\mathcal{A}\mathcal{T}}{0}_\gamma$. All three these terms are of order $\mathcal{O}(\Delta)$ and so according to the Zassenhaus formula we may write
\begin{equation}
	e^{\mathcal{T}^{-1}\mathcal{A}\mathcal{T}}=e^{\mathcal{A}_L}e^{\mathcal{A}_C}e^{\mathcal{A}_R}e^{\mathcal{O}(\Delta^2)}.
\end{equation}
Substituting this back into \eqref{Gfactor2} then produces
\begin{equation}
	{\textstyle {_\eta}{\matens{0}{e^{\mathcal{A}}e^{-it \tilde{H}_0}}{0}}{_\eta}\approx e^{{_\gamma}{\matens{0}{{\mathcal{T}^{-1}\mathcal{A}\mathcal{T}}}{0}}{_\gamma}}\prod_{k>0}\left(U_k^2+V_k^2 e^{-2it\tilde{\epsilon}_k}\right)}
	\label{vev1}
\end{equation}
which is again correct up to linear order in $\Delta$ in the exponentials' arguments. The remaining vacuum expectation value can be calculated by applying Wick's theorem on the level of the transformed operators $\bar{\gamma}^{(\dag)}_k=\mathcal{T}^{-1}\gamma^{(\dag)}_k\mathcal{T}$. The non-zero contractions are
\begin{align}
	{_\gamma}{\matens{0}{\bar{\gamma}_k\bar{\gamma}^\dag_k}{0}}{_\gamma}&=Q_k e^{2it\epst_k} 	&{_\gamma}{\matens{0}{\bar{\gamma}^\dag_k\bar{\gamma}_k}{0}}{_\gamma}&=\Lambda_k^2 Q_k\\
	{_\gamma}{\matens{0}{\bar{\gamma}^\dag_k\bar{\gamma}^\dag_{-k}}{0}}{_\gamma}&=i\Lambda_k Q_k e^{2it\epst_k} &{_\gamma}{\matens{0}{\bar{\gamma}_k\bar{\gamma}_{-k}}{0}}{_\gamma}&=i\Lambda_k Q_k
\end{align}
Combining these expressions with \eqref{Jexpression} and using $\Lambda_{-k}=-\Lambda_{k}$, $Q_{-k}=Q_{k}$ and $\epst_{-k}=\epst_{k}$ leads to
\begin{equation}
	{_\gamma}{\matens{0}{{\mathcal{T}^{-1}\mathcal{A}\mathcal{T}}}{0}}{_\gamma}=\frac{\Delta}{L}\sum_{k,k'}{}^{\boldsymbol{'}} Q_k Q_{k'} \Lambda_k \Lambda_{k'}M_{k,k'}
	\label{vevfinal}
\end{equation}
where		
\begin{align}
	M_{k,k'}=&\frac{4(e^{2it\epst_{k'}}-1)(e^{2it\epst_{k}}-\Lambda_k^2)K_1(k,k')}{\epst_{k'} \Lambda_k}+\frac{(e^{2it (\epst_k+\epst_{k'})}-1)(\cos(k+k')-2K_2(k,k'))}{\epst_k+\epst_{k'}}\nonumber\\
	&+\frac{(e^{2it \epst_{k'}}-e^{2it \epst_{k}})(\cos(k+k')+2K_2(k,k'))}{\epst_k-\epst_{k'}}
\end{align}
and
\begin{align}
K_1(k,k')=\sin[k+k'+2\theta_k(g_1)+2\theta_{k'}(g_1)]\sin^2[(k-k')/2]\\
K_2(k,k')=\cos[k+k'+2\theta_k(g_1)+2\theta_{k'}(g_1)]\sin^2[(k-k')/2].
\label{K12definition}
\end{align}
The primed summation in \eqref{vevfinal} excludes terms for which $k=\pm k'$. Combining \eqref{gfactorise}, \eqref{vev1} and \eqref{vevfinal} yields the final form of the return amplitude as
\begin{equation}
		G(t)\approx\prod_{k,k'}\left[1+(e^{-itD_{k,k'}}-1)\ave{\gamma^\dag_{k'}\gamma_{k'}}\ave{\gamma^\dag_{k}\gamma_{k}}\right]\prod_{k>0}\left(U_k^2+V_k^2 e^{-2it\tilde{\epsilon}_k}\right)\exp\left[\frac{\Delta}{L}\sum_{k,k'}{}^{\boldsymbol{'}} Q_k Q_{k'} \Lambda_k \Lambda_{k'}M_{k,k'}\right].
		\label{Gfinal}
\end{equation}
\subsection{Rate Function}
\label{sectionRF}
Starting from expression \eqref{Gfinal} we now proceed to calculate the corresponding rate function
\begin{equation}
	l(\Delta,t)=-\lim_{L\rightarrow\infty}\frac{1}{L}\log|G(t)|^2=-\lim_{L\rightarrow\infty}\frac{2}{L}{\mathrm{Re}}[\log G(t)].
	\label{ratefunction2}
\end{equation}
First consider the double product in $G(t)$ as it appears in \eqref{Gfinal}. The fact that $D_{k,k'}=\oforder{L^{-1}}$ allows the corresponding contribution to $l(t)$ to be written as 
\begin{equation}
	\lim_{L\rightarrow\infty}{\mathrm{Re}}\left[\frac{2it}{L}\sum_{k,k'}D_{k,k'} \ave{\gamma^\dag_{k'}\gamma_{k'}}\ave{\gamma^\dag_{k}\gamma_{k}}\right].
\end{equation}
Upon setting $\ave{\gamma^\dag_k \gamma_k}=\Lambda_k^2 Q_k$ and using $\Lambda_{-k}=-\Lambda_{k}$, $Q_{-k}=Q_{k}$ and $\epst_{-k}=\epst_{k}$ this expression becomes
\begin{equation}
	\lim_{L\rightarrow\infty}{\mathrm{Re}}\left[\frac{-16it\Delta}{L^2}\sum_{k,k'}K_2(k,k') \Lambda_k^2 \Lambda_{k'}^2 Q_k Q_{k'}\right].
\end{equation} 
Finally, combining the above with \eqref{Gfinal} yields
\begin{align}
	l(\Delta,t)=&-2\int_0^\pi\frac{dk}{2\pi}\ln|U_k^2+V_k^2 e^{-2it\epst_k}|\nonumber\\
	&-2\Delta{\mathrm{Re}}\left[\int_{-\pi}^\pi\frac{dk dk'}{(2\pi)^2} Q_k Q_{k'} \Lambda_k \Lambda_{k'} \left[M_{k,k'}+8it\Lambda_k\Lambda_{k'}K_2(k,k')\right]\right]+\oforder{\Delta^2}
	\label{lffinal}
\end{align} 
where the modified single particle energies are
\begin{equation}
	\epst_k=\epsilon_k+8\Delta\int_{-\pi}^{\pi}\frac{dk'}{2\pi}K_2(k,k').
\end{equation}
It will be useful to identify the linear order term in the expansion $l(\Delta,t)=l(0,t)+\Delta\, l^{(1)}(t)+\oforder{\Delta^2}$. To do so we expand the first term in \eqref{lffinal} to linear order in $\Delta$ (which enters through $\epst_k$) and replace $\epst_k\rightarrow\epsilon_k$ in the second term. This leads to 
\begin{align}
	\hspace{-0.3cm} l^{(1)}(t)=-2\,{\mathrm{Re}}\int_{-\pi}^\pi\frac{dk dk'}{(2\pi)^2}\left[ Q_k Q_{k'} \Lambda_k \Lambda_{k'}\left[M_{k,k'}+8it\Lambda_k\Lambda_{k'}K_2(k,k')\right]-8it\Lambda^2_kQ_k K_2(k,k')\right]
	\label{lf1}
\end{align} 
with all occurrences of $\epst_k$ replaced by $\epsilon_k$. For small $\Delta$ and short times the difference between $l(t)$ in \eqref{lffinal} and the truncated form $l(\Delta,t)=l(0,t)+\Delta\, l^{(1)}(t)$ is negligible. However, the truncation introduces secular terms and so \eqref{lffinal} remains more appropriate for the description of the dynamics at long times for which $t\sim \Delta^{-1}$. See Ref. \onlinecite{hackl_2009} for a detailed discussion of this point. We remark that at this stage it is not obvious how the perturbed critical times can be extracted from the results in \eqref{lffinal} or \eqref{lf1}. Certainly, no simple analytic solution is apparent. In fact, as shown in the next section, the truncation of $l(\Delta,t)$ at linear order introduces discontinuities (in time) which are not present in the exact result. Furthermore, the locations of these discontinuities do not coincide with the perturbed critical times. Despite these apparent difficulties it is still possible to extract both the shifts in the critical times and the change in the shapes of the cusps in $l(\Delta,t)$ from the perturbative results. The procedure for doing so is detailed in section \ref{sectionAnalyse}.
\section{Comparison to Numeric Results}
\begin{figure}[htbp]
\hspace{-0.5cm}
\begin{minipage}[b]{0.47\linewidth} \centering \includegraphics[height=5.7cm]{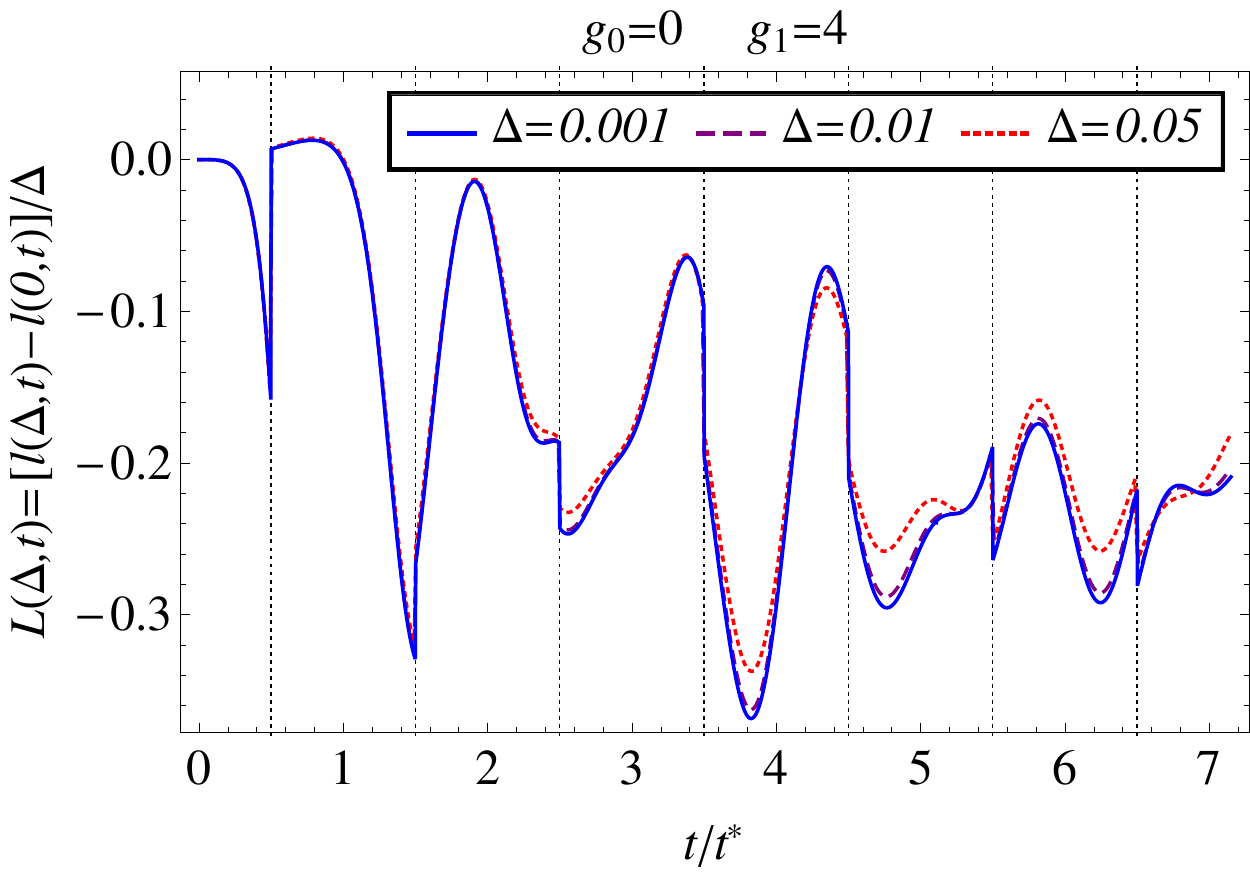} \end{minipage}
\hspace{0.5cm}
\begin{minipage}[b]{0.47\linewidth} \centering \includegraphics[height=5.7cm]{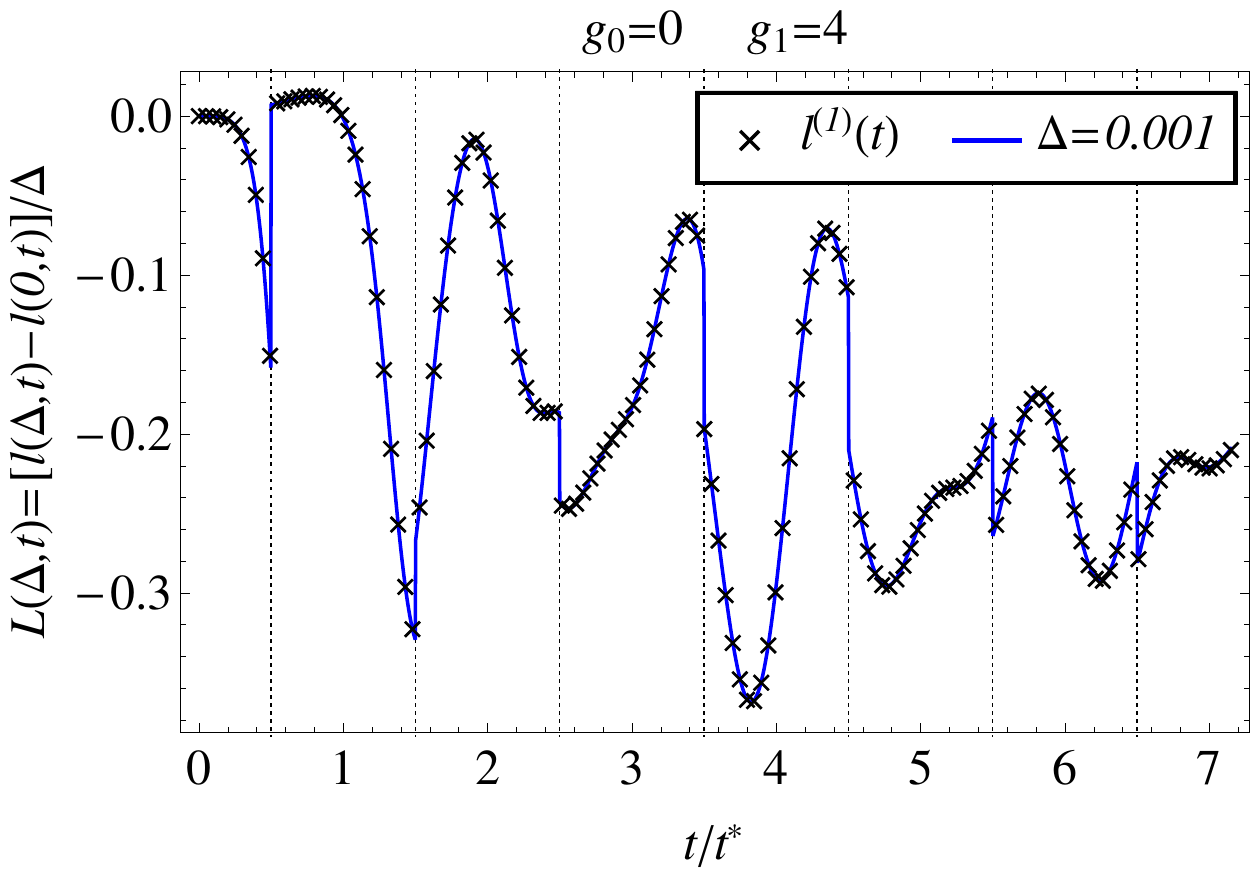} \end{minipage}
\caption{Results for the quench from $g_0=0$ to $g_1=4$ and non-zero $\Delta$. Left: tDMRG results for $L(\Delta,t)=(l(\Delta,t)-l(0,t))/\Delta$. Right: A comparison of $l^{1}(t)$ in \eqref{lf1} to the tDMRG estimate for $\Delta=0.001$. Vertical dashed lines indicate the unperturbed critical times $t^*_n$.}
\label{figureFMtoPM}
\end{figure}
\begin{figure}[htbp]
\hspace{-0.4cm}
  \includegraphics[height=5.7cm]{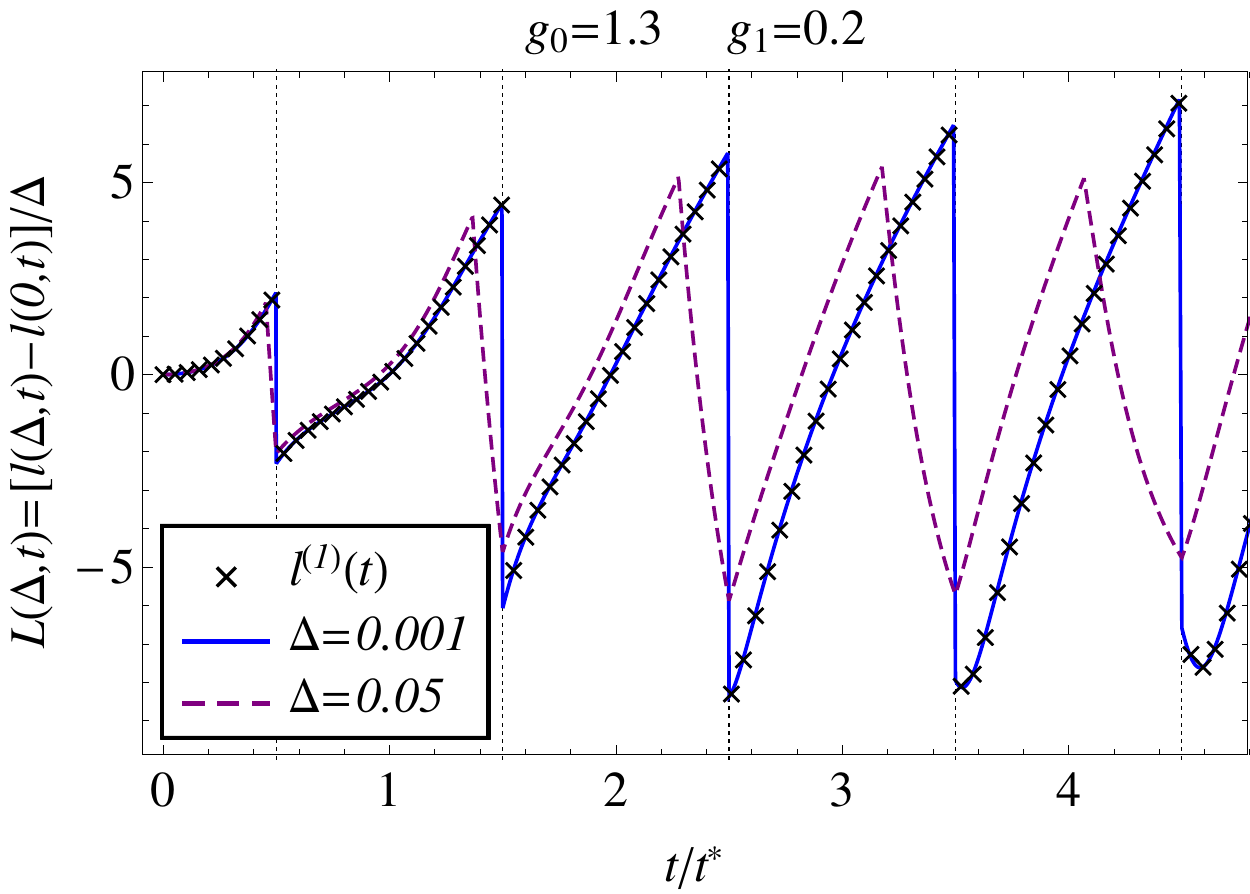}
	\caption{Results for the quench from $g_0=1.3$ to $g_1=0.2$ and non-zero $\Delta$. The prediction of $l^{1}(t)$ in \eqref{lf1} is compared to the tDMRG estimate $L(\Delta,t)=(l(\Delta,t)-l(0,t))/\Delta$. Vertical dashed lines indicate the unperturbed critical times $t^*_n$.}
	\label{figurePMtoFM}
\end{figure}

\label{sectionCompareDMRG}
To benchmark the perturbative calculation we have performed comparisons with results obtained using the time-dependent density matrix renormalisation group (tDMRG) algorithm. These numeric calculations are carried out directly in the thermodynamic limit; see Ref. \onlinecite{karrasch_2013} for details and further applications to this and related spin models. At weak coupling we expect the NNN interaction to perturb the rate function $l(\Delta,t)$ only slightly. Instead of considering $l(\Delta,t)$ itself, it is therefore more sensible to investigate $L(\Delta,t)=(l(\Delta,t)-l(0,t))/\Delta$. For times and couplings within the perturbative regime we expect $L(\Delta,t)$ to be well approximated by $l^{(1)}(t)$ in \eqref{lf1}. For a first comparison we consider a quench from the FM to PM phase with $g_0=0$ and $g_1=4$. The tDMRG results for several values of $\Delta$ are shown in Figure \ref{figureFMtoPM}.  The rate function itself appears in Figure \ref{figureratefunctions} and is clearly continuous at the critical times. The same holds for $L(\Delta,t)$, but it is found to vary very rapidly close to the critical times for small $\Delta$. On the horizontal scale of Figure \ref{figureFMtoPM} this appears as apparent discontinuities. We see that up to the seventh critical time the curves for $\Delta=0.01$ and $\Delta=0.001$ are almost indistinguishable. At these times and for $\Delta\lessapprox 0.01$ the linear order contribution to $l(\Delta,t)$ therefore dominates and we expect $l^{(1)}(t)$ and $L(\Delta,t)$ to be approximately equal. This is indeed the case, as can be seen in Figure \ref{figureFMtoPM}. We also note that, unlike $L(\Delta,t)$, $l^{(1)}(t)$ exhibits true discontinuities at the \emph{unperturbed} critical times $t^*_n$. This can be attributed to the divergence of the $\left.Q_{k^*}\right|_{\Delta=0}$ factors in \eqref{lf1} which occur at $t=t^*_n$ when $k=k^*$ with $\cos k^*=(1+g_0 g_1)/(g_0+g_1)$.\\

Figure \eqref{figurePMtoFM} shows the same comparison for a quench from the PM to the FM phase with $g_0=1.3$ and $g_1=0.2$. We again observe excellent agreement between the predictions of $l^{(1)}(t)$ in \eqref{lf1} and the tDMRG results for small $\Delta$. In this case $\Delta=0.05$ represents a strong NNN coupling which produces a large shift in the critical times. This results in the appearance of two sets of cusps in $L(\Delta,t)$ corresponding to cusps at the perturbed and unperturbed critical times present in $l(\Delta,t)$ and $l(0,t)$ respectively. This is a non-perturbative feature which cannot be reproduced at any finite order of perturbation theory. At first sight this might appear to prohibit the calculation the shifted critical times from the truncated form of the rate function $l(\Delta,t)\approx l(0,t)+\Delta l^{(1)}(t)$, as the latter only exhibits non-analyticies at the unperturbed critical times. In the next section we show that this is not the case, and that it is indeed possible to extract the linear order shifts in the critical times from our perturbative results. 

\subsection{Analysis of Non-analyticities}
\label{sectionAnalyse}
\begin{table}
\begin{minipage}[t]{0.47\linewidth} \renewcommand{\arraystretch}{1}
\renewcommand\tabcolsep{8pt}
\begin{tabular}{clll}
\hline
\hline
\multicolumn{2}{c}{} & \multicolumn{2}{c}{DMRG}\\
\cline{3-4}
\multicolumn{1}{c}{n} & \multicolumn{1}{c}{$\tilde{t}^*_n$} & \multicolumn{1}{c}{$\Delta=0.001$} & \multicolumn{1}{c}{$\Delta=0.005$} \\
\hline
\hline
0 & 0.01264 &  0.01256 & 0.01265 \\
1 & 0.009865 &  0.009906 & 0.009946\\
2 & -0.01862 &  -0.01855 & -0.01824\\
3 & -0.05783 &  -0.05771 & -0.05666\\
4 & -0.08834 &  -0.08797 & -0.08644\\
\hline
\hline
\end{tabular} \end{minipage}
\hspace{0.7cm}
\begin{minipage}[t]{0.47\linewidth} \renewcommand{\arraystretch}{1.0}
\renewcommand\tabcolsep{8pt}
\begin{tabular}{clll}
\hline
\hline
\multicolumn{2}{c}{} & \multicolumn{2}{c}{DMRG}\\
\cline{3-4}
\multicolumn{1}{c}{n} & \multicolumn{1}{c}{$\tilde{t}^*_n$} & \multicolumn{1}{c}{$\Delta=0.001$} & \multicolumn{1}{c}{$\Delta=0.005$} \\
\hline
\hline
0 & -1.870 & -1.866 & -1.853\\
1 & -5.513 & -5.497 & -5.434\\
2 &-10.30 & -10.25 & -10.05\\
3 & -14.90 & -14.86 & -14.69\\
4 & -19.75 & -19.66 & -19.36\\
\hline
\hline
\end{tabular} \vspace{0.0cm} \end{minipage}
\caption{The linear order shifts in the critical times due to the NNN interaction. The left (right) table shows results for the quench $g_0=0$ to $g_1=4$ ($g_0=1.3$ to $g_0=0.2$). The numerical tDMRG estimate $(t^*_{n,\Delta}-t^*_{n})/\Delta$ is shown for comparison.}
\label{tableshifts}
\end{table}
The cusps appearing in the return probability rate function are signatures of dynamical phase transitions in the post-quench dynamics. Here we analyse how the location and shape of these non-analyticities are affected by the perturbing NNN interaction. To this end it is useful to first return to the integrable case with $\Delta=0$, i.e. the transverse field Ising model, and consider two limiting examples which provide insight into the nature of these structures \cite{heyl_2013,karrasch_2013}. Consider a quench from $g_0=\infty$ to $g_1=0$. The rate function, for $L$ divisible by four, is then \mbox{$l(t,L)=-2\ln[\cos^L(t)+\sin^L(t)]/L$}. As $L\rightarrow\infty$ the value of $l(t,L)$ is determined by the largest term in the argument of the logarithm. In fact, in the thermodynamic limit $l(t)=\min\{f_1(t),f_2(t)\}$ with $f_1(t)=-\ln[\cos^2(t)]$ and $f_2(t)=-\ln[\sin^2(t)]$. This illustrates that the critical times $t^*_n=\pi/2(n+1/2)$ are not non-analytic points of $f_1(t)$ or $f_2(t)$ individually, but rather those times at which the two functions intersect and $l(t)$ switches between them. A similar picture emerges for the reverse FM to PM quench with $g_0=0$ and $g_1=\infty$, except here $f_1(t)$ and $f_2(t)$ have the additional interpretation of being the rate functions for transitions between different magnetisation sectors \cite{heyl_2013}. The tDMRG results shown in Figure \ref{figureratefunctions} suggest that this picture captures the generic nature of these non-analyticities for quenches across the critical point with finite $g_{0,1}$ and $\Delta$ as well.\\

We now consider a generic quench across the phase transition from $g_0$ to $g_1$ with \mbox{$\Delta\neq 0$}. Due to the NNN interaction the critical times will be shifted from $t^*_n$ in \eqref{H0criticaltimes} to $t^*_{n,\Delta}$. Based on the discussion above we assume that in a neighbourhood of each $t^*_{n,\Delta}$ there exist functions $f_{L,R}(\Delta,t)$, depending analytically on $t$ and $\Delta$, which form the left and right sides of the cusp. To be precise, $l(\Delta,t)=f_L(\Delta,t)$ for $t\leq t_{n,\Delta}^*$ and $l(\Delta,t)=f_R(\Delta,t)$ for $t\geq t_{n,\Delta}^*$. The particular critical time then satisfies $f_L(\Delta,t_{n,\Delta}^*)=f_R(\Delta,t_{n,\Delta}^*)$. To linear order in the coupling $\Delta$ we write $f_{L,R}(\Delta,t)=f^{(0)}_{L,R}(t)+\Delta f^{(1)}_{L,R}(t)$ and $t^*_{n,\Delta}=t^*_n+\Delta \tilde{t}^*_n$ where $f^{(0)}_{L}(t^*_n)=f^{(0)}_{R}(t^*_n)$. From this we can solve for $\tilde{t}^*_n$, which determines the leading order shift in the critical time, to find
\begin{equation}
	\tilde{t}^*_n=\frac{f^{(1)}_{R}(t^*_n)-f^{(1)}_{L}(t^*_n)}{\dot{f}^{(0)}_{L}(t^*_n)-\dot{f}^{(0)}_{R}(t^*_n)}.
	\label{ttilde}
\end{equation}
This expression can be evaluated using the analytic results for $l^{(0,1)}(t)$ by setting 
\begin{align}
	f^{(1)}_{L}(t^*_n)=\lim_{t\nearrow t^*_n} l^{(1)}(t)\hspace{2cm} \dot{f}^{(0)}_{L}(t^*_n)=\lim_{t\nearrow t^*_n} \dot{l}^{(0)}(t)\\
	f^{(1)}_{R}(t^*_n)=\lim_{t\searrow t^*_n} l^{(1)}(t)\hspace{2cm} \dot{f}^{(0)}_{R}(t^*_n)=\lim_{t\searrow t^*_n} \dot{l}^{(0)}(t)  
\end{align}
Table \ref{tableshifts} shows the results of this calculation together with the tDMRG estimate $(t^*_{n,\Delta}-t^*_{n})/\Delta$ and we again observe good agreement within the perturbative regime for both types of quenches.\\

As noted in Ref. \onlinecite{karrasch_2013} the NNN interaction appears to shift the critical times away from their periodic values at $\Delta=0$. Here we see that this is already a linear order effect. We have calculated $\tilde{t}^*_n$ up to $n=40$ but found no simple limiting behaviour. In particular, this non-periodicity rules out the possibility of accounting for the NNN interaction through a modified set of single particle energies in \eqref{H0results}. To quantify the change in the shape of the cusp we analyse the discontinuity in the first derivative of $l(t)$. Let \mbox{$\delta \dot{l}(\Delta,t^*_{n,\Delta})=\lim_{\epsilon\rightarrow0}[\dot{l}(\Delta,t^*_{n,\Delta}+\epsilon)-\dot{l}(\Delta,t^*_{n,\Delta}-\epsilon)]$} denote the jump in $\dot{l}(\Delta,t)$ at the critical time $t^*_{n,\Delta}$. To leading order we find
\begin{equation}
	\delta \dot{l}(\Delta,t^*_{n,\Delta})-\delta \dot{l}(0,t^*_{n})=\Delta\left[\dot{f}^{(1)}_R(t^*_n)-\dot{f}^{(1)}_L(t^*_n)+\tilde{t}^*_n[\ddot{f}^{(0)}_R(t^*_n)-\ddot{f}^{(0)}_L(t^*_n)]\right].
	\label{cuspchange}
\end{equation}
Estimates for this quantity can also be extracted from the tDMRG data.  We again find that these numeric estimates match the predictions of \eqref{cuspchange} very well, with a level of agreement similar to that seen in Table \ref{tableshifts}.

\section{Conclusions}
\label{sectionConclusion}
We have investigated the effect of the non-integrable next-nearest-neighbour (NNN) interaction on dynamical quantum phase transitions in the post-quench dynamics of the ANNNI model. This was done within a perturbative analytic framework based on the continuous unitary transformation approach to time evolution. These phase transitions manifest as cusps in the rate function of the return amplitude at a set of critical times. We have presented analytic results for the change in the shape and location of these cusps due to the perturbing NNN interaction. Our results support those of earlier numerical studies \cite{karrasch_2013} which demonstrated that these non-analytic features are robust with respect to the inclusion of the NNN interaction and depend on the coupling strength in a regular, thought complicated, way. In particular, we find that the shift of the critical times away from periodicity is already a linear order effect in the NNN coupling. 

\begin{acknowledgments}
JNK gratefully acknowledges the hospitality of the Institute for Theoretical Physics at the University of G\"{o}ttingen and the financial support of the HB \& MJ Thom trust. CK acknowledges the support of the Nanostructured Thermoelectrics program of LBNL. SK acknowledges support through SFB 1073 of the Deutsche Forschungsgemeinschaft (DFG).
\end{acknowledgments}

\appendix*
\section{}
\label{appendixA}
Here we summarise the derivation of expressions \eqref{H0fermionic} and \eqref{H1fermionic} for $H_0$ and $H_1$ and provide expressions for the coefficients appearing in the latter. First we apply the Wigner-Jordan transformation $\sigma_i^x=1-2c_i^\dag c_i$ and $\sigma_i^z=\prod_{j<i}(2c^\dag_jc_j-1)(c_i^\dag+c_i)$ to $H=H_0+H_1$ in \eqref{fullH} to obtain
\begin{align}
	H_0&=\sum_i(c_i-c_i^\dag)(c_{i+1}+c_{i+1}^\dag)+g\sum_i(2c_i^\dag c_i-1)\\
	H_1&=\Delta\sum_i(c_i-c^\dag_i)(1-2c_{i+1}^\dag c_{i+1})(c^\dag_{i+2}+c_{i+2})
\end{align}
where periodic (antiperiodic) boundary conditions are enforced in the odd (even) particle number sector. Fourier transforming to $c_k=L^{-1/2}\sum_j c_j e^{-ikj}$ then produces
\begin{align}
	H_0&=\sum_k\left[2(g-\cos(k))c_k^\dag c_k+i\sin(k)(c_{-k}^\dag c_k^\dag+c_{-k} c_k)-g\right]\\
	H_1&=-\Delta\left[H_{1,1}+H_{1,2}+H_{1,3}\right]
\end{align}
where
\begin{align}
	H_{1,1}&=\sum_k 2\cos(2k) c_k^\dag c_k-i\sin(2k)(c_{-k}^\dag c_k^\dag+c_{-k}c_k)\\
	H_{1,2}&=\frac{4}{L}\sum_{\mathbf{k}}\delta(k_1+k_2-k_3-k_4)\cos(k_2+k_4)c^\dag_{k_1}c^\dag_{k_2}c_{k_3}c_{k_4}\\
	H_{1,3}&=-\frac{2}{L}\sum_{\mathbf{k}}\delta(k_1+k_2+k_3-k_4)\left[e^{i(k_2-k_3)}c^\dag_{k_1}c^\dag_{k_2}c^\dag_{k_3}c_{k_4}+h.c.\right].
\end{align}
In the odd (even) sector $k$ is quantized in integer (half-integer) multiples of $2\pi/L$. Finally we introduce the Bogoliubov fermions $\gamma^{(\dag)}_k$ by $c_k=u_k\gamma_k+i v_k\gamma^\dag_{-k}$ where $u_k=\cos(\theta_k)$ and $v_k=\sin(\theta_k)$ with $\tan(2\theta_k)=\sin(k)/(g_1-\cos(k))$. Solutions to the latter equation are chosen such that $\theta_k\in[0,\pi/2]$ for $k\in[0,\pi]$ and $\theta_k\in[-\pi/2,0]$ when $k\in[-\pi,0)$. To handle the lengthy algebra resulting from the Bogoliubov transformation we used the SNEG package \cite{zitko_2011} for \textsc{Mathematica} to extract the coefficients in \eqref{H1fermionic}. We find that
\begin{equation}
	B(k)=\frac{8\Delta}{L}\sum_{k'}K_2(k,k')\mtext{1}{and}C(k)=\frac{4i\Delta}{L}\sum_{k'}K_1(k,k')
\end{equation}
with $K_{1,2}$ given in \eqref{K12definition}. In terms of the three auxiliary functions 
\begin{align}
	D'=&\left[u_{k_1} u_{k_2} u_{k_2'} v_{-k_1'}(\sin (k_1-k_2)-2 \sin (k_1'+k_1))\right.\nonumber\\
	&+u_{k_2} v_{-k_1} u_{k_2'} v_{-k_1'} (\cos (k_1'+k_1)-\cos(k_1-k_2))\\
	&\left.+u_{k_1} u_{k_2} u_{k_1'} u_{k_2'} \cos (k_2'+k_1)\right]+(k_i\leftrightarrow k_i')\nonumber\\
	\nonumber\\
	E'=&u_{k_3} v_{-k_1} (v_{-k_2}u_{k_1'}\sin (k_1-k_2)+2 u_{k_2} v_{-k_1'} \sin (k_1'+k_2))\nonumber\\
    &+2 u_{k_2}u_{k_3} v_{-k_1} u_{k_1'} (\cos (k_1'+k_2)-\cos (k_1-k_2))\\
		&+u_{k_1}u_{k_2}u_{k_3}u_{k_1'}\sin (k_1-k_2)\nonumber\\
		\nonumber\\
	F'=&u_{k_3} u_{k_4} v_{-k_1}(u_{k_2} \sin(k_2-k_3)+v_{-k_2} \cos(k_1-k_3))
\end{align}
the remaining coefficients read
\begin{align}
D(k_1',k_2',k_1,k_2)=&\delta_{k_1+k_2,k_1'+k_2'}\frac{2\Delta}{L}\left[D'+(u_k\rightarrow -v_{-k},v_k\rightarrow u_{-k})\right]\\
E(k_1,k_2,k_3,k_1')=&\delta_{k_1+k_2+k_3,k_1'}\frac{2i\Delta}{L}\left[E'-(u_k\rightarrow -v_{-k},v_k\rightarrow u_{-k})\right]\\
F(k_1,k_2,k_3,k_4)=&\delta_{k_1+k_2+k_3+k_4}\frac{2\Delta}{L}\left[F'+(u_k\rightarrow -v_{-k},v_k\rightarrow u_{-k})\right].
\end{align}


\begin{thebibliography}{29}%
\makeatletter
\providecommand \@ifxundefined [1]{%
 \@ifx{#1\undefined}
}%
\providecommand \@ifnum [1]{%
 \ifnum #1\expandafter \@firstoftwo
 \else \expandafter \@secondoftwo
 \fi
}%
\providecommand \@ifx [1]{%
 \ifx #1\expandafter \@firstoftwo
 \else \expandafter \@secondoftwo
 \fi
}%
\providecommand \natexlab [1]{#1}%
\providecommand \enquote  [1]{``#1''}%
\providecommand \bibnamefont  [1]{#1}%
\providecommand \bibfnamefont [1]{#1}%
\providecommand \citenamefont [1]{#1}%
\providecommand \href@noop [0]{\@secondoftwo}%
\providecommand \href [0]{\begingroup \@sanitize@url \@href}%
\providecommand \@href[1]{\@@startlink{#1}\@@href}%
\providecommand \@@href[1]{\endgroup#1\@@endlink}%
\providecommand \@sanitize@url [0]{\catcode `\\12\catcode `\$12\catcode
  `\&12\catcode `\#12\catcode `\^12\catcode `\_12\catcode `\%12\relax}%
\providecommand \@@startlink[1]{}%
\providecommand \@@endlink[0]{}%
\providecommand \url  [0]{\begingroup\@sanitize@url \@url }%
\providecommand \@url [1]{\endgroup\@href {#1}{\urlprefix }}%
\providecommand \urlprefix  [0]{URL }%
\providecommand \Eprint [0]{\href }%
\providecommand \doibase [0]{http://dx.doi.org/}%
\providecommand \selectlanguage [0]{\@gobble}%
\providecommand \bibinfo  [0]{\@secondoftwo}%
\providecommand \bibfield  [0]{\@secondoftwo}%
\providecommand \translation [1]{[#1]}%
\providecommand \BibitemOpen [0]{}%
\providecommand \bibitemStop [0]{}%
\providecommand \bibitemNoStop [0]{.\EOS\space}%
\providecommand \EOS [0]{\spacefactor3000\relax}%
\providecommand \BibitemShut  [1]{\csname bibitem#1\endcsname}%
\let\auto@bib@innerbib\@empty
\bibitem [{\citenamefont {Greiner}\ \emph {et~al.}(2002)\citenamefont
  {Greiner}, \citenamefont {M{\"a}ndel}, \citenamefont {Hansch},\ and\
  \citenamefont {Bloch}}]{greiner_2002}%
  \BibitemOpen
  \bibfield  {author} {\bibinfo {author} {\bibfnamefont {M.}~\bibnamefont
  {Greiner}}, \bibinfo {author} {\bibfnamefont {O.}~\bibnamefont {M{\"a}ndel}},
  \bibinfo {author} {\bibfnamefont {T.~W.}\ \bibnamefont {Hansch}}, \ and\
  \bibinfo {author} {\bibfnamefont {I.}~\bibnamefont {Bloch}},\ }\href
  {\doibase 10.1038/nature00968} {\bibfield  {journal} {\bibinfo  {journal}
  {Nature}\ }\textbf {\bibinfo {volume} {419}},\ \bibinfo {pages} {51}
  (\bibinfo {year} {2002})}\BibitemShut {NoStop}%
\bibitem [{\citenamefont {Kinoshita}\ \emph {et~al.}(2006)\citenamefont
  {Kinoshita}, \citenamefont {Wenger},\ and\ \citenamefont
  {Weiss}}]{kinoshita_2006}%
  \BibitemOpen
  \bibfield  {author} {\bibinfo {author} {\bibfnamefont {T.}~\bibnamefont
  {Kinoshita}}, \bibinfo {author} {\bibfnamefont {T.}~\bibnamefont {Wenger}}, \
  and\ \bibinfo {author} {\bibfnamefont {D.~S.}\ \bibnamefont {Weiss}},\ }\href
  {\doibase 10.1038/nature04693} {\bibfield  {journal} {\bibinfo  {journal}
  {Nature}\ }\textbf {\bibinfo {volume} {440}},\ \bibinfo {pages} {900}
  (\bibinfo {year} {2006})}\BibitemShut {NoStop}%
\bibitem [{\citenamefont {Polkovnikov}\ \emph {et~al.}(2011)\citenamefont
  {Polkovnikov}, \citenamefont {Sengupta}, \citenamefont {Silva},\ and\
  \citenamefont {Vengalattore}}]{polkovnikov_2011}%
  \BibitemOpen
  \bibfield  {author} {\bibinfo {author} {\bibfnamefont {A.}~\bibnamefont
  {Polkovnikov}}, \bibinfo {author} {\bibfnamefont {K.}~\bibnamefont
  {Sengupta}}, \bibinfo {author} {\bibfnamefont {A.}~\bibnamefont {Silva}}, \
  and\ \bibinfo {author} {\bibfnamefont {M.}~\bibnamefont {Vengalattore}},\
  }\href {\doibase 10.1103/RevModPhys.83.863} {\bibfield  {journal} {\bibinfo
  {journal} {Rev. Mod. Phys.}\ }\textbf {\bibinfo {volume} {83}},\ \bibinfo
  {pages} {863} (\bibinfo {year} {2011})}\BibitemShut {NoStop}%
\bibitem [{\citenamefont {Heyl}\ \emph {et~al.}(2013)\citenamefont {Heyl},
  \citenamefont {Polkovnikov},\ and\ \citenamefont {Kehrein}}]{heyl_2013}%
  \BibitemOpen
  \bibfield  {author} {\bibinfo {author} {\bibfnamefont {M.}~\bibnamefont
  {Heyl}}, \bibinfo {author} {\bibfnamefont {A.}~\bibnamefont {Polkovnikov}}, \
  and\ \bibinfo {author} {\bibfnamefont {S.}~\bibnamefont {Kehrein}},\ }\href
  {\doibase 10.1103/PhysRevLett.110.135704} {\bibfield  {journal} {\bibinfo
  {journal} {Phys. Rev. Lett.}\ }\textbf {\bibinfo {volume} {110}},\ \bibinfo
  {pages} {135704} (\bibinfo {year} {2013})}\BibitemShut {NoStop}%
\bibitem [{\citenamefont {Fisher}(1965)}]{fisher_1965}%
  \BibitemOpen
  \bibfield  {author} {\bibinfo {author} {\bibfnamefont {M.}~\bibnamefont
  {Fisher}},\ }in\ \href@noop {} {\emph {\bibinfo {booktitle} {Lectures in
  Theoretical Physics {VII}}}}\ (\bibinfo  {publisher} {University of Colorado
  Press},\ \bibinfo {address} {Boulder, Colorado},\ \bibinfo {year}
  {1965})\BibitemShut {NoStop}%
\bibitem [{\citenamefont {Gong}\ and\ \citenamefont {Duan}(2013)}]{gong_2013}%
  \BibitemOpen
  \bibfield  {author} {\bibinfo {author} {\bibfnamefont {Z.-X.}\ \bibnamefont
  {Gong}}\ and\ \bibinfo {author} {\bibfnamefont {L.-M.}\ \bibnamefont
  {Duan}},\ }\href {\doibase 10.1088/1367-2630/15/11/113051} {\bibfield
  {journal} {\bibinfo  {journal} {New J. Phys.}\ }\textbf {\bibinfo {volume}
  {15}},\ \bibinfo {pages} {113051} (\bibinfo {year} {2013})}\BibitemShut
  {NoStop}%
\bibitem [{\citenamefont {Fagotti}(2013)}]{fagotti_2013}%
  \BibitemOpen
  \bibfield  {author} {\bibinfo {author} {\bibfnamefont {M.}~\bibnamefont
  {Fagotti}},\ }\href {http://arxiv.org/abs/1308.0277} {\bibfield  {journal}
  {\bibinfo  {journal} {{arXiv}:1308.0277}\ } (\bibinfo {year}
  {2013})}\BibitemShut {NoStop}%
\bibitem [{\citenamefont {Heyl}(2014)}]{heyl_2014}%
  \BibitemOpen
  \bibfield  {author} {\bibinfo {author} {\bibfnamefont {M.}~\bibnamefont
  {Heyl}},\ }\href {http://arxiv.org/abs/1403.4570} {\bibfield  {journal}
  {\bibinfo  {journal} {{arXiv}:1403.4570}\ } (\bibinfo {year}
  {2014})}\BibitemShut {NoStop}%
\bibitem [{\citenamefont {Vajna}\ and\ \citenamefont
  {D{\'o}ra}(2014)}]{vajna_2014}%
  \BibitemOpen
  \bibfield  {author} {\bibinfo {author} {\bibfnamefont {S.}~\bibnamefont
  {Vajna}}\ and\ \bibinfo {author} {\bibfnamefont {B.}~\bibnamefont
  {D{\'o}ra}},\ }\href {\doibase 10.1103/PhysRevB.89.161105} {\bibfield
  {journal} {\bibinfo  {journal} {Phys. Rev. B}\ }\textbf {\bibinfo {volume}
  {89}},\ \bibinfo {pages} {161105} (\bibinfo {year} {2014})}\BibitemShut
  {NoStop}%
\bibitem [{\citenamefont {Andraschko}\ and\ \citenamefont
  {Sirker}(2014)}]{andraschko_2014}%
  \BibitemOpen
  \bibfield  {author} {\bibinfo {author} {\bibfnamefont {F.}~\bibnamefont
  {Andraschko}}\ and\ \bibinfo {author} {\bibfnamefont {J.}~\bibnamefont
  {Sirker}},\ }\href {\doibase 10.1103/PhysRevB.89.125120} {\bibfield
  {journal} {\bibinfo  {journal} {Phys. Rev. B}\ }\textbf {\bibinfo {volume}
  {89}},\ \bibinfo {pages} {125120} (\bibinfo {year} {2014})}\BibitemShut
  {NoStop}%
\bibitem [{\citenamefont {Hickey}\ \emph {et~al.}(2014)\citenamefont {Hickey},
  \citenamefont {Genway},\ and\ \citenamefont {Garrahan}}]{hickey_2014}%
  \BibitemOpen
  \bibfield  {author} {\bibinfo {author} {\bibfnamefont {J.~M.}\ \bibnamefont
  {Hickey}}, \bibinfo {author} {\bibfnamefont {S.}~\bibnamefont {Genway}}, \
  and\ \bibinfo {author} {\bibfnamefont {J.~P.}\ \bibnamefont {Garrahan}},\
  }\href {\doibase 10.1103/PhysRevB.89.054301} {\bibfield  {journal} {\bibinfo
  {journal} {Phys. Rev. B}\ }\textbf {\bibinfo {volume} {89}},\ \bibinfo
  {pages} {054301} (\bibinfo {year} {2014})}\BibitemShut {NoStop}%
\bibitem [{\citenamefont {Karrasch}\ and\ \citenamefont
  {Schuricht}(2013)}]{karrasch_2013}%
  \BibitemOpen
  \bibfield  {author} {\bibinfo {author} {\bibfnamefont {C.}~\bibnamefont
  {Karrasch}}\ and\ \bibinfo {author} {\bibfnamefont {D.}~\bibnamefont
  {Schuricht}},\ }\href {\doibase 10.1103/PhysRevB.87.195104} {\bibfield
  {journal} {\bibinfo  {journal} {Phys. Rev. B}\ }\textbf {\bibinfo {volume}
  {87}},\ \bibinfo {pages} {195104} (\bibinfo {year} {2013})}\BibitemShut
  {NoStop}%
\bibitem [{\citenamefont {Selke}(1988)}]{selke_1988}%
  \BibitemOpen
  \bibfield  {author} {\bibinfo {author} {\bibfnamefont {W.}~\bibnamefont
  {Selke}},\ }\href {\doibase 10.1016/0370-1573(88)90140-8} {\bibfield
  {journal} {\bibinfo  {journal} {Phys. Rep.}\ }\textbf {\bibinfo {volume}
  {170}},\ \bibinfo {pages} {213} (\bibinfo {year} {1988})}\BibitemShut
  {NoStop}%
\bibitem [{\citenamefont {Suzuki}\ \emph {et~al.}(2013)\citenamefont {Suzuki},
  \citenamefont {Inoue},\ and\ \citenamefont {Chakrabarti}}]{suzuki_2013}%
  \BibitemOpen
  \bibfield  {author} {\bibinfo {author} {\bibfnamefont {S.}~\bibnamefont
  {Suzuki}}, \bibinfo {author} {\bibfnamefont {J.-i.}\ \bibnamefont {Inoue}}, \
  and\ \bibinfo {author} {\bibfnamefont {B.~K.}\ \bibnamefont {Chakrabarti}},\
  }\href@noop {} {\emph {\bibinfo {title} {Quantum Ising phases and transitions
  in transverse Ising models}}}\ (\bibinfo  {publisher} {Springer},\ \bibinfo
  {address} {Heidelberg},\ \bibinfo {year} {2013})\BibitemShut {NoStop}%
\bibitem [{\citenamefont {Silva}(2008)}]{silva_2008}%
  \BibitemOpen
  \bibfield  {author} {\bibinfo {author} {\bibfnamefont {A.}~\bibnamefont
  {Silva}},\ }\href {\doibase 10.1103/PhysRevLett.101.120603} {\bibfield
  {journal} {\bibinfo  {journal} {Phys. Rev. Lett.}\ }\textbf {\bibinfo
  {volume} {101}},\ \bibinfo {pages} {120603} (\bibinfo {year}
  {2008})}\BibitemShut {NoStop}%
\bibitem [{\citenamefont {Sachdev}(2011)}]{sachdev_2011}%
  \BibitemOpen
  \bibfield  {author} {\bibinfo {author} {\bibfnamefont {S.}~\bibnamefont
  {Sachdev}},\ }\href@noop {} {\emph {\bibinfo {title} {Quantum Phase
  Transitions}}}\ (\bibinfo  {publisher} {Cambridge University Press},\
  \bibinfo {year} {2011})\BibitemShut {NoStop}%
\bibitem [{\citenamefont {Pollmann}\ \emph {et~al.}(2010)\citenamefont
  {Pollmann}, \citenamefont {Mukerjee}, \citenamefont {Green},\ and\
  \citenamefont {Moore}}]{pollmann_2010}%
  \BibitemOpen
  \bibfield  {author} {\bibinfo {author} {\bibfnamefont {F.}~\bibnamefont
  {Pollmann}}, \bibinfo {author} {\bibfnamefont {S.}~\bibnamefont {Mukerjee}},
  \bibinfo {author} {\bibfnamefont {A.~G.}\ \bibnamefont {Green}}, \ and\
  \bibinfo {author} {\bibfnamefont {J.~E.}\ \bibnamefont {Moore}},\ }\href
  {\doibase 10.1103/PhysRevE.81.020101} {\bibfield  {journal} {\bibinfo
  {journal} {Phys. Rev. E}\ }\textbf {\bibinfo {volume} {81}},\ \bibinfo
  {pages} {020101} (\bibinfo {year} {2010})}\BibitemShut {NoStop}%
\bibitem [{\citenamefont {Calabrese}\ \emph {et~al.}(2011)\citenamefont
  {Calabrese}, \citenamefont {Essler},\ and\ \citenamefont
  {Fagotti}}]{calabrese_2011}%
  \BibitemOpen
  \bibfield  {author} {\bibinfo {author} {\bibfnamefont {P.}~\bibnamefont
  {Calabrese}}, \bibinfo {author} {\bibfnamefont {F.~H.~L.}\ \bibnamefont
  {Essler}}, \ and\ \bibinfo {author} {\bibfnamefont {M.}~\bibnamefont
  {Fagotti}},\ }\href {\doibase 10.1103/PhysRevLett.106.227203} {\bibfield
  {journal} {\bibinfo  {journal} {Phys. Rev. Lett.}\ }\textbf {\bibinfo
  {volume} {106}},\ \bibinfo {pages} {227203} (\bibinfo {year}
  {2011})}\BibitemShut {NoStop}%
\bibitem [{\citenamefont {Calabrese}\ \emph {et~al.}(2012)\citenamefont
  {Calabrese}, \citenamefont {Essler},\ and\ \citenamefont
  {Fagotti}}]{calabrese_2012}%
  \BibitemOpen
  \bibfield  {author} {\bibinfo {author} {\bibfnamefont {P.}~\bibnamefont
  {Calabrese}}, \bibinfo {author} {\bibfnamefont {F.~H.~L.}\ \bibnamefont
  {Essler}}, \ and\ \bibinfo {author} {\bibfnamefont {M.}~\bibnamefont
  {Fagotti}},\ }\href {\doibase 10.1088/1742-5468/2012/07/P07016} {\bibfield
  {journal} {\bibinfo  {journal} {J. Stat. Mech.}\ }\textbf {\bibinfo {volume}
  {2012}},\ \bibinfo {pages} {P07016} (\bibinfo {year} {2012})}\BibitemShut
  {NoStop}%
\bibitem [{\citenamefont {Hackl}\ and\ \citenamefont
  {Kehrein}(2009)}]{hackl_2009}%
  \BibitemOpen
  \bibfield  {author} {\bibinfo {author} {\bibfnamefont {A.}~\bibnamefont
  {Hackl}}\ and\ \bibinfo {author} {\bibfnamefont {S.}~\bibnamefont
  {Kehrein}},\ }\href {\doibase 10.1088/0953-8984/21/1/015601} {\bibfield
  {journal} {\bibinfo  {journal} {J. Phys.: Condens. Matter}\ }\textbf
  {\bibinfo {volume} {21}},\ \bibinfo {pages} {015601} (\bibinfo {year}
  {2009})}\BibitemShut {NoStop}%
\bibitem [{\citenamefont {Wegner}(1994)}]{wegner_1994}%
  \BibitemOpen
  \bibfield  {author} {\bibinfo {author} {\bibfnamefont {F.}~\bibnamefont
  {Wegner}},\ }\href {\doibase 10.1002/andp.19945060203} {\bibfield  {journal}
  {\bibinfo  {journal} {Ann. Phys.}\ }\textbf {\bibinfo {volume} {506}},\
  \bibinfo {pages} {77} (\bibinfo {year} {1994})}\BibitemShut {NoStop}%
\bibitem [{\citenamefont {Kehrein}(2006)}]{kehrein_2006}%
  \BibitemOpen
  \bibfield  {author} {\bibinfo {author} {\bibfnamefont {S.}~\bibnamefont
  {Kehrein}},\ }\href@noop {} {\emph {\bibinfo {title} {The flow equation
  approach to many-particle systems}}}\ (\bibinfo  {publisher} {Springer},\
  \bibinfo {address} {Berlin},\ \bibinfo {year} {2006})\BibitemShut {NoStop}%
\bibitem [{\citenamefont {Kehrein}(2005)}]{kehrein_2005}%
  \BibitemOpen
  \bibfield  {author} {\bibinfo {author} {\bibfnamefont {S.}~\bibnamefont
  {Kehrein}},\ }\href {\doibase 10.1103/PhysRevLett.95.056602} {\bibfield
  {journal} {\bibinfo  {journal} {Phys. Rev. Lett.}\ }\textbf {\bibinfo
  {volume} {95}},\ \bibinfo {pages} {056602} (\bibinfo {year}
  {2005})}\BibitemShut {NoStop}%
\bibitem [{\citenamefont {Hackl}\ and\ \citenamefont
  {Kehrein}(2008)}]{hackl_2008}%
  \BibitemOpen
  \bibfield  {author} {\bibinfo {author} {\bibfnamefont {A.}~\bibnamefont
  {Hackl}}\ and\ \bibinfo {author} {\bibfnamefont {S.}~\bibnamefont
  {Kehrein}},\ }\href {\doibase 10.1103/PhysRevB.78.092303} {\bibfield
  {journal} {\bibinfo  {journal} {Phys. Rev. B}\ }\textbf {\bibinfo {volume}
  {78}},\ \bibinfo {pages} {092303} (\bibinfo {year} {2008})}\BibitemShut
  {NoStop}%
\bibitem [{\citenamefont {Moeckel}\ and\ \citenamefont
  {Kehrein}(2008)}]{moeckel_2008}%
  \BibitemOpen
  \bibfield  {author} {\bibinfo {author} {\bibfnamefont {M.}~\bibnamefont
  {Moeckel}}\ and\ \bibinfo {author} {\bibfnamefont {S.}~\bibnamefont
  {Kehrein}},\ }\href {\doibase 10.1103/PhysRevLett.100.175702} {\bibfield
  {journal} {\bibinfo  {journal} {Phys. Rev. Lett.}\ }\textbf {\bibinfo
  {volume} {100}},\ \bibinfo {pages} {175702} (\bibinfo {year}
  {2008})}\BibitemShut {NoStop}%
\bibitem [{\citenamefont {Moeckel}\ and\ \citenamefont
  {Kehrein}(2009)}]{moeckel_2009}%
  \BibitemOpen
  \bibfield  {author} {\bibinfo {author} {\bibfnamefont {M.}~\bibnamefont
  {Moeckel}}\ and\ \bibinfo {author} {\bibfnamefont {S.}~\bibnamefont
  {Kehrein}},\ }\href {\doibase 10.1016/j.aop.2009.03.009} {\bibfield
  {journal} {\bibinfo  {journal} {Ann. Phys.}\ }\textbf {\bibinfo {volume}
  {324}},\ \bibinfo {pages} {2146} (\bibinfo {year} {2009})}\BibitemShut
  {NoStop}%
\bibitem [{\citenamefont {Moeckel}\ and\ \citenamefont
  {Kehrein}(2010)}]{moeckel_2010}%
  \BibitemOpen
  \bibfield  {author} {\bibinfo {author} {\bibfnamefont {M.}~\bibnamefont
  {Moeckel}}\ and\ \bibinfo {author} {\bibfnamefont {S.}~\bibnamefont
  {Kehrein}},\ }\href {\doibase 10.1088/1367-2630/12/5/055016} {\bibfield
  {journal} {\bibinfo  {journal} {New J. Phys.}\ }\textbf {\bibinfo {volume}
  {12}},\ \bibinfo {pages} {055016} (\bibinfo {year} {2010})}\BibitemShut
  {NoStop}%
\bibitem [{\citenamefont {Essler}\ \emph {et~al.}(2014)\citenamefont {Essler},
  \citenamefont {Kehrein}, \citenamefont {Manmana},\ and\ \citenamefont
  {Robinson}}]{essler_2014}%
  \BibitemOpen
  \bibfield  {author} {\bibinfo {author} {\bibfnamefont {F.~H.~L.}\
  \bibnamefont {Essler}}, \bibinfo {author} {\bibfnamefont {S.}~\bibnamefont
  {Kehrein}}, \bibinfo {author} {\bibfnamefont {S.~R.}\ \bibnamefont
  {Manmana}}, \ and\ \bibinfo {author} {\bibfnamefont {N.~J.}\ \bibnamefont
  {Robinson}},\ }\href {\doibase 10.1103/PhysRevB.89.165104} {\bibfield
  {journal} {\bibinfo  {journal} {Phys. Rev. B}\ }\textbf {\bibinfo {volume}
  {89}},\ \bibinfo {pages} {165104} (\bibinfo {year} {2014})}\BibitemShut
  {NoStop}%
\bibitem [{\citenamefont {{\v Z}itko}(2011)}]{zitko_2011}%
  \BibitemOpen
  \bibfield  {author} {\bibinfo {author} {\bibfnamefont {R.}~\bibnamefont {{\v
  Z}itko}},\ }\href {\doibase 10.1016/j.cpc.2011.05.013} {\bibfield  {journal}
  {\bibinfo  {journal} {Comput. Phys. Commun.}\ }\textbf {\bibinfo {volume}
  {182}},\ \bibinfo {pages} {2259} (\bibinfo {year} {2011})}\BibitemShut
  {NoStop}%
\end{thebibliography}
%
\end{document}